\documentclass{article}
\usepackage{color}
\usepackage{amssymb}
\usepackage{amsmath}
\usepackage{latexsym}
\usepackage{slashed, amsthm}
\usepackage{graphicx}

\title{The modular Dirac equation}
\author{C. Rugina$^{1}$
	\\
	\\ \small{Department of Physics}
	\\	\small{University of Bucharest, Bucharest, Romania}
	\\}

\begin{document}
	
	\maketitle 
	{\let\thefootnote\relax\footnotetext{{\em $^{1}$Email}:
			christina.rugina11@alumni.imperial.ac.uk}}
	
	\begin{abstract}
		
		\noindent
		We introduce a new equation we dubbed the modular Dirac equation to see and reconstruct a spin 1/2 particle at the center of a nearly $AdS_2$ spacetime in the entanglement wedge reconstruction paradigm and we study hidden symmetries of this spacetime, too. Various properties of the Dirac modular operator are studied: a generalized Tomita-Takesaki construction, the connection to the Schwarzian derivative of the logarithm of the modular Dirac operator, the link with the an allowable complex metric, the connection to regenesis, we write the corresponding Lagrangian of the modular Dirac equation and we put in perspective some limitations of the current bulk reconstruction in the case of unknown couplings.
	\end{abstract}

\section{Introduction}

\noindent
AdS/CFT provided us with an answer to avoid the black hole information paradox \cite{fuzz1}, but also led us in the direction of bulk reconstruction, building a bulk-boundary dictionary, which approximately allows the reconstruction of bulk operators. Both the solution and the initial issue, the information paradox, are powerful paradigms that taught us many things, for instance the existence of structure at the horizon \cite{BHC1,EPR, softphoton} and of black hole microstates, which in string theory reproduce the Bekenstein-Hawking entropy of the black hole \cite{twocharge1,twocharge2, threecharge}. In this context, we reach out here to study aspects of entanglement wedge reconstruction (EWR), as a larger bulk reconstruction avenue than causal wedge. As we evade the information paradox, we see that's possible because information is encoded in the Hawking radiation especially after Page time. The existence of the Page curve \cite{P9} of the entanglement entropy of the black hole is a response to the firewall paradox \cite{AMPS} and it is a proof of black hole unitary evaporation and a consequence of the fact that the interior of the black hole is encoded in the outgoing, thermal Hawking radiation. 

\medskip
\noindent
One other important development in EWR is the link with the quantum information aspect of quantum error correcting codes (QEC), which actually relies on perturbatively extending the quantum HRT formula in 1/N expansion\cite{H27, H56}. A constructive approach of EWR as QEC is carried out in \cite{Harlow}. There is considerable effort in EWR to be able to give information about interior operators, starting with a given black hole microstate, like in the Kourkoulou-Maldacena construction \cite{KM} or state-dependent developments as in \cite{P32,EWR_see_13}. To avoid a firewall then a reduced number of states, the code subspace is proposed\cite{Almheiri2018}. Avoiding a firewall is also a cornerstone of EWR and this work was extensively carried out in the literature - see as a starting point of the discussion \cite{AMPS, EPR}.

\medskip
\noindent
We place our work in this diverse context and we try here to see and actively reconstruct a bulk spin 1/2 particle at the center of a $AdS_2$ spacetime with an equation that gives the evolution of this bulk spin connected via a wormhole in the embedding spacetime to the boundary. We employ methods from raising a closed operator to a complex power to solve this evolution equation we dubbed the modular Dirac equation and we take a look at hidden symmetries generated by it in the nearly $AdS_2$ spacetime. We also investigate a number of properties of the associated modular Dirac operator ranging from a generalized Tomita-Takesaki construction to the implications of the existence of a complex metric saddle associated with the $AdS_2$ spacetime, the connection to the Schwarzian theory and regenesis. We also show that there are some limitations to this bulk reconstruction method in the case of unknown couplings.

\section{Solving the modular equation}

\medskip
\noindent
Taking into account the EWR paradigm, we are now ready to tackle the problem of reconstructing the spin at the center of the bulk of an AdS spacetime using a novel technique, which aligns well with EWR. Using concepts from modular flow and Connes co-cycle flow, we will introduce a new equation which describes the evolution of a bulk spin 1/2 connected via a wormhole with the boundary.

\medskip
\noindent
The $\text{AdS}_2$ metric in higher dimensional embedding coordinates $(Y^+, Y^-, Y_{-1})$ can be expressed as \cite{Maldacena2017}:

\begin{equation}
ds^2 = d Y^+ dY^- -(dY_{-1})^2
\end{equation}

\medskip
\noindent
In bulk coordinates \cite{Maldacena2017} these embedding coordinates write:

\begin{equation}
Y^+ = \sinh \rho  e^t \hspace{0.5cm} Y^- = - \sinh \rho  e^{-t} \hspace{0.5cm} Y_{-1} = \cosh \rho
\end{equation}

\medskip
\noindent
and in these bulk coordinates the metric is:

\begin{equation}
ds^2 = d \rho^2 - \sinh ^2 \rho dt^2
\end{equation}

\medskip
\noindent
Also the Connes co-cycle can also be defined as \cite{EWR_see, Bousso}:

\begin{equation}
(\mathcal{D} \phi: \mathcal{D} \psi)_s \equiv \Delta^{is}_{k|\phi} \Delta ^{-is}_{k| \psi} = \Delta^{is}_{\phi} \Delta ^{-is}_{\phi| \psi}
\end{equation}

\medskip
\noindent
Note that the relative modular operator $\Delta$ can be written as a function of density matrices and the algebras $\mathcal{A}_A, \mathcal{A}'_A$ associated with a region A of the spacetime as \cite{EWR_see}:

\begin{equation}
\Delta_{\psi | \phi, \mathcal{A}_A} = \rho_{\psi; \mathcal{A}'_A} \otimes \rho ^{-1}_{\phi; \mathcal{A}_A}
\end{equation}

\medskip
\noindent
So the Connes co-cycle flow is in fact introduced on two disjoint open spacetime areas (A,A') with two corresponding algebras $\mathcal{A}, \mathcal{A'}$ (much like in Tomita-Takesaki theory which we will briefly discuss later), which can be assimilated with the left and right sides in a two-sided black hole, for instance, as a particular case.
Consequently, given the above, we can define in the embedding of the $\text{AdS}_2$ spacetime a modular Dirac equation which describes the evolution of spin 1/2 bulk particles, connected via a wormhole to the boundary. We will give a justification of why this equations takes this form later in the paper. To be noted that the embedding spacetime is needed since the topology contains a wormhole:

\begin{equation}
(\cosh x_{-1} (\overset{\leftrightarrow} {D^{1+is}})- m)\psi_s = 0
\end{equation}

\medskip
\noindent
where D is the usual Dirac operator in $\text{AdS}_2$ spacetime, and

\begin{equation}
\alpha \overset{\leftrightarrow} D \beta = \alpha D \beta - \beta D \alpha
\end{equation}

\medskip
\noindent
and $\psi_s$ is a family of spinors such that:

\begin{equation}
(\mathcal{D} \psi: \mathcal{D} \psi)_s | \psi\rangle = | \psi_s \rangle
\end{equation}

\medskip
\noindent
where s is playing the role of modular time. Here and in the relation above $\psi$ is a cyclic and separating spinor and $|\psi_s\rangle$ is a Connes co-cycle flowed family of spinors parameterized by the modular time. The modular equation reduces to the regular Dirac equation in the limit of compactified dimension  $x_{-1} \rightarrow 0$ and s = 0, so in $\text{AdS}_2$ spacetime with zero modular time. So the wormhole that forms away from the limit $x_{-1} \rightarrow 0$ propagates fast to the boundary of the AdS spacetime by analogy with the propagation of the bubble of nothing in Kaluza-Klein theories noted by Witten in \cite{Witten} and this provides an exact reconstruction of the spin at the center of the bulk up to unknown couplings, as we will see in the last subsection of section 4 of this paper. To be noted that there is a connection between our work here and that of \cite{EWR_see} in that both papers propose 'seeing' and acting on the bulk which is part of the the entanglement wedge. We take things a step further here and in $AdS_2$ we propose a mechanism via the above evolution equation to 'see' and act on deep bulk, even at the center of it. 

\medskip
\noindent
 Also $ \overset{\leftrightarrow} {D^{1+is}}$ is part of a Guillemin algebra \cite{Guillemin} such that:

\begin{equation} \label{Gui}
D^{z}= q(a(z)r^{mz}) + R(z)
\end{equation}

\medskip
\noindent
The Dirac operator has the usual form:

\begin{equation}
\gamma^a D_a = \gamma^a e_a +\frac{1}{4} \gamma^a \gamma^b \gamma^c \omega_{bc} (e_a)
\end{equation}

\noindent
where $\omega$ is the usual spin connection. Note that in the coordinates given,

\begin{equation}
e^0= d\rho \hspace{1.0cm} e^1 = \sinh \rho dt
\end{equation}

\noindent
and the corresponding gamma matrices can be written as a function of Pauli matrices as:

\begin{equation}
\gamma^0 = \sigma_1, \hspace{1.0cm} \gamma^1 = -i \sigma_2.
\end{equation}

\noindent
We use the following equation to get the spin connections:

\begin{equation}
d e^a +{\omega^a}_b \wedge e^b =0
\end{equation}

\noindent
Consequently the Dirac operator can be written as:

\begin{equation}
D =\left[ \sigma_1 \partial_{\rho} + \frac{1}{2}\left( \sigma_3 \cosh \rho -i \sigma_2 \sinh \rho \right) \partial_t \right]
\end{equation}

\noindent
Based on the work on Dirac operators raised at complex powers in \cite{ComplexOp}, we find that the solution to the modular Dirac equation is:

\begin{equation}
\psi_s (\rho, t, x_{-1}) = \cosh x_{-1} E_{1+is} (\rho +i t)
\end{equation}

\noindent
where $E_{1+is}$ is obtained from the solution in \cite{ComplexOp} setting $\alpha =0$, $\mu = 1+is$, $m= \mu$,  $\xi=1$ and $\epsilon = \sinh \rho$. So:

\begin{multline}
E_\mu (\rho) = \frac{1+ e^{-i\pi \mu}}{2} \frac{e^{-i\pi}}{2^{\mu -1} \pi^{(\mu -1)/2}}\frac{\Gamma(\frac{1}{2})}{\Gamma(\frac{1}{2}\mu)} D_0^{1/2} (\rho) - \\ \\
		-\frac{1-e^{i\pi \mu}}{2} \frac{e^{-i\pi(m-\mu)/2}}{2^{\mu-1} \pi^{(m-1)/2}} \frac{1}{\Gamma(\frac{1}{2}(1+\mu))} \left( D^1_{-1} ( \rho) - \sinh \rho D_0^1 (\rho) \right)
\end{multline}

\noindent
where D is the Gegenbauer function defined as \cite{Abramovitz}:

\begin{equation}
D^\lambda_\nu (z) = \frac{ 2^{(1-2\lambda)} \sqrt{\pi} \Gamma (\nu + 2 \lambda)}{\nu! \Gamma(\lambda)} {}_2F_1 (-\nu;\nu+2 \lambda;\lambda +\frac{1}{2}; \frac{1-z}{2})
\end{equation}

\noindent
To be noted that this solution has a well-posed initial value problem in conformity with \cite{ComplexOp} and this normalization is realized on a unit hyperbolic ball, also according to \cite{ComplexOp}. Even if $AdS_2$ is not a globally hyperbolic spacetime, the above hyperbolic solution gives a sense in which the evolution of a bulk spin in $AdS_2$ is known within a certain degree of approximation even in the case of unknown couplings.

\section{Hidden modular symmetries}

\noindent
We now turn to investigate here hidden symmetries of the $AdS_2$ spacetime, which posses a rich modular structure. We introduce a modular closed conformal Killing-Yanos tensor (MCCKY) in analogy with its CCKY counterpart and the equation writes:

\begin{equation}
\nabla^{1+is}_a k_{bc} = \nabla^{1+is}_{[a} k_{bc]} + 2 g_{a[b} \tilde{k}_{c]}
\end{equation}

\noindent
where

\begin{equation}
\tilde{k}_c = \nabla^{1+is}_b k^b_c
\end{equation}

\medskip
\noindent
Note that the MCCKY equation reduces to the CCKY equation for s=0. Also note that $k_{00} = k_{11}$, so the only not null components are $k_{01}= - k_{10}$, then the MCCKY equation reduces to:

\begin{equation}\label{CCKY}
\nabla_0^{1+is} k_{01} =0
\end{equation}

\medskip
\noindent
The solution to this equation is given by \cite{ComplexOp}:

\begin{equation}
k_{01} = \frac{1+ e^{-i\pi (1+is)}}{2} Z_{1+is} + \frac{1- e^{-i\pi(1+is)}}{2is} \rho Z_{is}
\end{equation}

\noindent
where

\begin{equation}
Z_\mu = \frac{\rho^{\mu -3}}{\pi^{1/2} 2^{\mu-1} \Gamma(\frac{1}{2}\mu) \Gamma(\frac{1}{2}(\mu-1))}
\end{equation}

\medskip
\noindent
The modular KY tensor is actually a scalar given by:

\begin{equation}
f = *k
\end{equation}

\medskip
\noindent
and the modular conformal SK tensor is given by:

\begin{equation}
S_{ac} = k_{ab} {k^b}_c.
\end{equation}

\medskip
\noindent
Moreover, we introduce the modular Dirac-type operator as:

\begin{equation}
D^{(1+is)}_t = i \gamma^{\mu_1} {k_{\mu_1}}^{\mu_2} \nabla^{1+is}_{\mu_2} -\frac{1}{6}\gamma^{\mu_1} \gamma^{\mu_2} \gamma^\mu \nabla^{1+is}_\mu k_{\mu_1 \mu_2}
\end{equation}

\medskip
\noindent
One can show that:

\begin{equation}
\{D_t^{1+is}, D^{1+is}\}=0.
\end{equation}

\medskip
\noindent
and this is straightforward given that the regular Dirac-type operator anticommutes with the Dirac operator according to for instance \cite{Visi}. Namely:

\begin{equation}
	\{D_t, D\}=0.
\end{equation}

\noindent
and so one just needs to prove that:

\begin{equation}
	\{D_t^{is}, D^{is}\}=0.
\end{equation}

\noindent
which can be done using the fact that $D^{is}$ obeys the Guillemin algebra and relation (\ref{Gui}).
One can immediately see that the following is true given that the only not-null components are $k_{01} = - k_{10}$:

\begin{equation}\label{UnitRoots}
k^\mu_\alpha k_{\mu \beta} = g_{\alpha \beta}.
\end{equation}

\medskip
\noindent
This last relation defines the modular unit roots of this spacetime. Also one can prove that if k is a modular unit root then:

\begin{equation}
D^2_t = D^2
\end{equation}

\medskip
\noindent
which follows along the same lines as with the proof in the case of regular Dirac-type operators, following for instance \cite{Visi}.
If we perform a Duval-Eisenhart lift of the $AdS_2$ to 4 dimensions, the lifted solutions k of the $(\ref{UnitRoots})$
form a basis of three modular unit roots complex structures to offer $AdS_4$ a so-called modular hyper-K\"{a}hler structure, which is analogous to a regular hyper-K\"{a}hler structure. To be noted that this 4-dimensional modular hyper-K\"{a}hler structure admits a dimensional reduced modular K\"{a}hler analogue in $AdS_2$.

\medskip
\noindent
The above lifted modular unit roots build a modular symplectic 2-form in 4 dimensions:

\begin{equation}
	\omega = \frac{1}{2} k_{\mu \nu} dx^\mu \wedge dx^\nu
\end{equation}

\medskip
\noindent
which is the modular symplectic form in the modular K\"{a}hler structure. So we define a modular K\"{a}hler structure in 4 dimensions (M, g, k, $\omega$), where (M,g) is our 4-dimensional manifold with the lifted metric, (M, $\omega$) is a modular symplectic manifold and (M,k) is a modular complex manifold with:

\begin{equation}
	g(X,Y)= g(kX, kY) \hspace{1.cm} \omega (X,Y) = g(kX,Y)
\end{equation}

\noindent
It follows that three modular K\"{a}hler structures correspond to a modular hyper-k\"{a}hler structure (M,g, $k^1,k^2,k^3, \omega^1, \omega^2, \omega^3$) with a quaternionic relation between $k^i$. Now this modular hyper-K\"{a}hler structure can be dimensionally reduced to a modular K\"{a}hler structure in $AdS_2$. 

\medskip
\noindent
The hidden symmetries presented in this section offer a glimpse into the hidden entanglement between the center of the bulk and the boundary of $AdS_2$, entanglement governed by the evolution equation dubbed the modular Dirac equation. These symmetries underline the fact that things are not what they seem in $AdS_2$ and that deep bulk reconstruction is possible in the most unfavorable situations even when couplings are unknown albeit in an approximate manner in this case. These symmetries also underline the fact that there is a subtle connection between the geometric aspects of this spacetime and the evolution operator presented here, this geometry being influenced in a hidden manner by the modular Dirac operator.

\section{Various properties of the modular Dirac operator}

\subsection{A Tomita-Takesaki construction}

\medskip
\noindent
We will write below a construction mirroring and generalizing the Tomita-Takesaki construction \cite{Papadodimas}:

\begin{equation}
	SA|\psi_{s,0} \rangle = A^\dagger | \psi_{s,0} \rangle
\end{equation}

\begin{equation}
	\Delta = S^\dagger S,\hspace{0.5cm} S= J \Delta^{1/2} \hspace{0.5cm} \tilde{D}_\omega = J \overset{\leftrightarrow} {D^{1+is +i2 \pi n}_\omega} J
\end{equation}

\noindent
where $A\in \mathcal{A}$ is a generalized type $I_\infty$ von Neumann algebra, as we will see below, $|\psi_{s,0} \rangle$ is the modular Dirac operator ground state wavefunction, S is an anti-linear operator related to the modular conjugation anti-unitary operator J and $\overset{\leftrightarrow} {D^{1+is +i2 \pi n}_\omega}$ is the Fourier transform of the modular Dirac operator:

\begin{equation}
	\overset{\leftrightarrow} {D^{1+is +i2 \pi n}_\omega} = \int_{-\infty}^{\infty} dt e^{i\omega t} \overset{\leftrightarrow} {D^{1+is +i2 \pi n}}
\end{equation}

\medskip
\noindent
The infinite-dimensional algebra $\mathcal{A}$ which is type $I_\infty$ generalized von Neumann is defined as:

\begin{equation}
	\mathcal{A} = span\{D_{\omega_1}, \cdots,  D_{\omega_n}, \cdots\}
\end{equation}

\noindent
where $D_{\omega_n} = \overset{\leftrightarrow} {D^{1+is +i2 \pi n}_\omega}$. This is a generalized von Neumann algebra in the sense that while the operators are bounded and they are defined on the Hilbert space generated by the eigenfunctions base, the generalized von Neumann algebra admits an infinity of commutants as below.Then there are in infinity of commutant algebras $\mathcal{A}'_n$ of type I generalized von Neumann defined as follows:

\begin{equation}
	\mathcal{A}'_n = span\{D_{\omega_1}, \cdots, D_{\omega_n}\}
\end{equation}

\noindent
In regular von Neumann algebras there is a bipartite division mirroring a thermofield double state, where the von Neumann algebra is causally complete and its commutant is the causal complement. In our situation we have multipartite entanglement, where n operators (universes) are entangled and then we can add another one to the entanglement web to make it n+1, until we reach the result that an infinity of universes (operators) are entangled. However, the set of n operators can be successively considered the causal complement of the infinite set. 
The multiverse modular Dirac wavefunction can also be written as:

\begin{multline}
	\psi_s^n (x) = \sum_{j=1}^2 \int\frac{d^2k}{(2 \pi)^2} \frac{1}{\sqrt{2 E_k}}[({a_{\vec{k}}^j})^{1+is+ 2 i\pi n} u^j(\vec{k}) e^{i\vec{k} \vec{x}} + \\ \\
	+ ({a_{\vec{k}}^j}^\dagger)^{1+is+ 2 i\pi n} v^j(\vec{k}) e^{-i\vec{k} \vec{x}}]
\end{multline}

\noindent
where $a_{\vec{k}}$, $a^\dagger_{\vec{k}}$ are regular creation and annihilation operators and $u^j(\vec{k})$, $v^j(\vec{k})$ are the regular Dirac spinors. That being said, going back to the algebras in our construction, they are analogous to the ones in the Tomita-Takesaki construction as we have seen above. The physical interpretation of the commutant $\mathcal{A}'_n$ is that they correspond to a multiverse, so we can say we have introduced a construction which generalizes the Tomita-Takesaki one to the multiverse situation.

\subsection{The connection of the modular Dirac operator to the Schwarzian derivative}

\medskip
\noindent
Let us first notice that since $\overset{\leftrightarrow} {D^{1+is}}$ is a bounded operator, then:

\begin{equation}
	\overset{\leftrightarrow} {D^{1+is}} = e^{(1+is) \ln \overset{\leftrightarrow} {D}}
\end{equation}

\medskip
\noindent
and that:

\begin{equation}
	\ln (1+ \overset{\leftrightarrow} {D} ) \simeq \overset{\leftrightarrow} {D}- \frac{\overset{\leftrightarrow} {D}^2}{2} + \cdots
\end{equation}

\medskip
\noindent
One can immediately see that the generalization of the Schwarzian derivative can be written as:

\begin{equation}
	\{\overset{\leftrightarrow} {D}, z\} = - \frac{\overset{\leftrightarrow} {D}^2}{2}
\end{equation}

\medskip
\noindent
It follows from here that:

\begin{equation}
	\ln (1+ \overset{\leftrightarrow} {D} ) \simeq \overset{\leftrightarrow} {D} - 	\{\overset{\leftrightarrow} {D}, z\} + \cdots
\end{equation}

\noindent
From here there is just one step to:

\begin{equation}\label{eqLN}
	\ln \overset{\leftrightarrow} {D} \simeq -1 + \overset{\leftrightarrow} {D} - 	\{\overset{\leftrightarrow} {D}, z\} + \cdots
\end{equation}

\medskip
\noindent
So we have proven that the logarithm of the modular Dirac operator $\overset{\leftrightarrow} {D}^{(1+is)}$ is in the second degree approximation the Schwarzian derivative and so there is a subtle connection between the modular Dirac Lagrangian and the Schwarzian action and theory. The modular Dirac Lagrangian can be written as:

\begin{equation}
	\mathcal{L} = i \bar{\psi} \cosh x_{-1} \overset{\leftrightarrow} {D^{(1+is)}} \psi - m \psi \bar{\psi}
\end{equation}

\medskip
\noindent
One can also write the allowable complex metric (in the sense of \cite{CWitten}) associated with the $AdS_2$ spacetime as:

\begin{equation}
	ds^2 = d \rho^{(2+2is)}  -(\sinh \rho)^{(2+2is)} dt^{(2+2is)}
\end{equation}

\medskip
\noindent
One can see that this metric is allowable by writing it as:

\begin{equation}
	d \rho^{(2+2is)} = e^{(2+2i) \ln d\rho}
\end{equation}

\noindent
and so forth with the rest of the metric terms. The complex metric has the physical interpretation of complex saddles and provides a regularization of the singular double cone metric. 

\medskip
\noindent
One can also introduce by analogy with Araki relative entropy, the fermionic relative entropy between bulk and boundary:

\begin{equation}
S_{rel Bb} = \langle \Psi | \log \overset{\leftrightarrow} {D^{(1+is)}} | \Psi \rangle
\end{equation}

\medskip
\noindent
One can see that this observable takes complex values, but one can prove that the real part of S is greater equal than zero and that due to the monotonicity of Re S then the second law of thermodynamics holds, in total analogy with the Araki entropy \cite{TTWitten}. The imaginary part of the entropy is linked here to the non-locality of the physical problem under description, see also \cite{Poland} for other uses of complex (Shannon) entropy.

\medskip
\noindent
Following \cite{TTWitten} one can prove that the real part of the entropy is greater equal than zero based on the fact that:

\begin{equation}
-\log \overset{\leftrightarrow} {D} \ge 1- \overset{\leftrightarrow} {D}
\end{equation} 

\noindent
where we introduce following \cite{TTWitten} the notion of inequality between bounded operators starting from the definition of positive operators - it is said that  $ P \ge 0$ if:

\begin{equation}
	\langle \chi | P | \chi \rangle \ge 0
\end{equation} 

\noindent
for any $| \chi \rangle$. Then along the same lines as in \cite{TTWitten} one can prove monotonicity for the real part (or imaginary part for that matter) of $S_{rel Bb}$ and so it is proven that the usual second law of thermodynamics holds for the real part of the entropy. However, the complex entropy obeys the complexified version of the second law of thermodynamics, much like the replica wormholes saddles. So if we consider $S= |S| e^{i\theta}$ , |S| is monotonic and always greater than zero.

\subsection{Regenesis and why the modular Dirac equation looks like it looks}

\noindent
Here we try to sketch a proof of why the modular Dirac equation looks the way it looks based on compatibility with large c CFT theory (chaotic theory) in a regenesis context- see \cite{HongLiu}. Keeping the notations in \cite{HongLiu}, we introduce that:

\begin{equation}\label{eqV}
	V = \cosh x_{-1} \overset{\leftrightarrow} {D^{(1+is)}}= \cosh x_{-1} e^{(1+is)(-1+ \overset{\leftrightarrow} {D} +\{\overset{\leftrightarrow} {D}, z\} + \cdots)},
\end{equation}

\noindent
which makes sense because that's the interaction operator in our modular Lagrangian. The left and right operators $\mathcal{O}_L$ and $\mathcal{O}_R$ define V as:

\begin{equation}
V = \frac{1}{L}\int_{-L/2}^{L/2} dx	\mathcal{O}_L \mathcal{O}_R.
\end{equation}	

\noindent
With these notations it follows from \cite{HongLiu} that the correlator over the thermofield double state equals:

\begin{equation}\label{eq1}
\langle \Psi_\beta|	\mathcal{O}_L(0, x_i) \mathcal{O}_R (0, x_j)| \Psi_\beta \rangle = \frac{G}{(2\cosh \frac{2 \pi}{\beta} x_{ij} + 2)^{\Delta_{\mathcal{O}}}}
\end{equation}

\noindent
In our case the correlator based on the full-blown Dirac modular operator in the large c CFT on the boundary, so describing an ensemble of chaotic spins 1/2 at the center of the bulk and in terms of $\psi_s$ is written as:

\begin{equation}
G_{CFT}(p) \sim \frac{1}{(\cosh x_{-1} p^{1+is} +1)^\Delta}
\end{equation}	

\noindent
And this propagator in terms of the first approximation expansion of $\overset{\leftrightarrow} {D}$ is after a conformal rescaling of $x_{-1}$:

\begin{equation}\label{eq2}
	G^0_{CFT} \sim \frac{1}{(\cosh x_{-1}  +1)^\Delta}
\end{equation}	

\noindent
Comparing $(\ref{eq1})$ with $(\ref{eq2})$ we see that if we relate $\psi_s$ with $\Psi_\beta$ the two expressions coincide ($\Delta$-s coincide, too, since we have that V is the interaction term in both theories). Note that the modular Hamiltonian commutes with the modular Dirac operator and then the corresponding eigenstates are $ | \bar{n}_L s\rangle$ and $|n_R s\rangle$:

\begin{equation}
	[H, \overset{\leftrightarrow} {D^{(1+is)}}]=0
\end{equation}

\noindent
In this eigenbasis the thermofield double state can also be decomposed as:

\begin{equation}
	| \Psi_\beta \rangle = \frac{1}{Z_\beta} \sum_n e^-{\frac{\beta E_{n n}}{2}} |\bar{n}_L n\rangle |n_R n \rangle
\end{equation}

\noindent 
and the solution to the modular Dirac equation writes:

\begin{equation}
	\psi_s = \sum_n c_{ns} |\bar{n}_L s \rangle | n_R s \rangle
\end{equation}

\noindent
and the $c_{ns}$ are coefficients which can be determined analytically.

\noindent
So the two expressions (\ref{eq1}) and (\ref{eq2}) coincide in the first order approximation, which gives us some confidence that we got the right description for the modular Dirac equation taking $x_{-1} = \frac{2 \pi}{\beta} x_{ij}$. Coming back to the regenesis phenomenon, keeping the notations in \cite{HongLiu} we conclude that if we have a signal at the center of the bulk with an island formed of chaotic 1/2 spins, then the signal re-emerges on the boundary after the scrambling time. The phenomenon of regenesis can be captured in a formula the following way \cite{HongLiu}:

\begin{equation}
\langle J^b (t_s, \vec{x})\rangle_g \approx C(g) \phi^R(-t_s, \vec{x})
\end{equation}

\noindent
where $J^b$ is the re-emergent boundary regenesis signal, $\phi^R$ is the orginal central bulk excitation. Note that:

\begin{equation}
	C(g) = - 2G_J \text{Im} \langle e^{-igV}\rangle
\end{equation}

\noindent
with V given by (\ref{eqV}). Moreover, the quantities (correlator) W and $B_0$ introduced in \cite{HongLiu} relevant for regenesis in the case of large c CFT  scale as (for spins 1/2 ensembles):

\begin{equation}
	B_0 \sim \frac{1}{c^2}, \hspace{1.cm} W \sim \frac{1}{c^3}
\end{equation}

\noindent
where c is the boundary CFT central charge.

\subsection{The effect of unknown couplings on the current proposed bulk reconstruction}

\noindent
We have so far not investigated the effect of the couplings of the modular Dirac operator on our proposed bulk reconstruction. Following \cite{AlmheiriLin} we notice that in the case of almost $\text{AdS}_2$ with the coupling $\chi$ in $\text{H}_{tot} = \text{H} + \chi(u) \text{V}$ we can recover their result for the effective matter action in the first degree approximation of V. This means that we can't fully reconstruct the bulk in the case of unknown couplings, we can do that only in the first degree approximation.

\medskip
\noindent
Let's start by taking a look at the matter action:

\begin{equation}
-I_m = \int dx_{-1} d \rho dt \left[ (\cosh x_{-1})^{\frac{1}{1+is}}  \overset{\leftrightarrow} {D} \right]^{1+is} \chi(u) \chi (u')
\end{equation}

\noindent
where as in \cite{AlmheiriLin} $\chi(u) = J$ for $0< u < \tau$  and $\chi(u) = J'$ for $\tau < u < \beta$. We recall from (\ref{eqLN}) that:

\begin{equation}
	\overset{\leftrightarrow} {D} = e^{[-1 + \overset{\leftrightarrow} {D} - 	\{\overset{\leftrightarrow} {D}, z\} + \cdots]}
\end{equation}

\noindent
With this in mind it follows that going to the complex plane:

\begin{equation}
	-I_m \sim \int dz [(\cos z)^{\frac{1}{1+is}}]^{(1+is)} \chi(u) \chi (u')
\end{equation}

\noindent
Performing a change of variables one gets:

\begin{equation}
-I_m \sim  \int dz \left[\frac{1}{\sin^2 z}\right]^{-1/2} \chi(u) \chi (u')
\end{equation}

\noindent
To be noted that in our case the exact proportionality factor in this integration doesn't matter that much since it can be absorbed into the unknown couplings. In the end we obtain a similar result with \cite{AlmheiriLin} and their conclusions hold for us, too:

\begin{equation}
-I_m \sim  \int du_1 du_2 \left[\frac{1}{\sin^2 \frac{(u_1- u_2)}{2}}\right]^{-1/2} \chi(u) \chi (u')
\end{equation}

\noindent
where one could assimilate $u_1 = u_R$ and $u_2 = u_L$ to be the left right coordinates on the ends of the wormhole.

\section{Conclusions}

\medskip
\noindent
We have presented here the solution and the hidden symmetries of the modular Dirac equation in nearly $AdS_2$ spacetime and its embedding which describe a 1/2 spin at the center of the $AdS_2$ spacetime connected via a wormhole with the boundary. We have also presented some properties of the modular Dirac operator. It turns out the equation and the solution make use of complex power operators as part of the Guillemin algebra, relying also on Connes co-cycle concepts. The solution of the equation is given by a superposition of Gegenbauer functions and hidden symmetries implies the existence of new modular closed conformal tensors and modular St\"{a}ckel-Killing tensors. Modular Dirac-type operators are written out and in the end a hidden symmetry of the nearly $AdS_2$ dubbed modular hyper-K\"{a}hler structure is retrieved. A generalized Tomita-Takesaki construction is introduced and also the connection with the Schwarzian theory is emphasized. A complex metric and the lagrangian corresponding to the modular Dirac operator are written out. We present a justification of the form of the modular Dirac operator based on regenesis, then we finally comment on how unknown couplings affect how realistic this bulk reconstruction actually is.

\medskip
\noindent
This paper brings new insight into seeing at the center of the $AdS_2$ spacetime as part of the EWR program and also subsequently transferring information from the center of the spacetime to the boundary and this is achieved by solving the modular Dirac equation. Further studies could encompass a more detailed account of the hidden structures of the nearly $AdS_2$ spacetime.

\section*{Acknowledgements}
We thank Virgil B\u{a}ran, Eliahu Cohen, Dana Ioan, C\u{a}lin L\u{a}z\u{a}roiu, Andy Ludu, Howard Schnitzer and Claude Warnick for encouragement, comments on the draft and helpful discussions.


\begin{thebibliography}{99}
	
		\bibitem{fuzz1}
	S. D. Mathur, "The information paradox: A pedagogical introduction", Class. Quant. Grav. \textbf{26}, 224001 (2009), arxiv:0909.1038.
	
	\bibitem{BHC1}
	L. Susskind, L. Thorlacius, J. Uglum, "The stretched horizon and black hole complementarity", Phys.Rev.D \textbf{48}, 3743 (1993), arxiv: hep-th/9306069.
	
	\bibitem{EPR}
	J. Maldacena, L. Susskind, "Cool horizons for entangled black holes", Fortsch. Phys. \textbf{61}, 781 (2013), arxiv: 1306.0533.
	
	\bibitem{softphoton}
	S. W. Hawking, M. J. Perry, A. Strominger, "Soft Hair on Black Holes", Phys. Rev. Lett. \textbf{116}, 231301 (2016), arxiv:1601.00921.
	
	\bibitem{twocharge1}
	A. Sen, "Extremal black holes and elementary string states", Mod. Phys. Lett. A \textbf{10}, 2081 (1995), hep-th/9504147.
	
	\bibitem{twocharge2}
	A. Dabholkar, "Exact counting of black hole microstates", Phys. Rev. Lett. \textbf{94} 241301, (2005), hep-th/0409148.
	
	
	\bibitem{threecharge}
	A. Strominger, C. Vafa, "Microscopic Origin of the Bekenstein-Hawking Entropy", Phys. Lett. B \textbf{379}, 99 (1996), hep-th/9601029.
	
	
	\bibitem{P9}
	D. Page, "Information in black hole radiation",  Phys. Rev. Lett. \textbf{71}, 3743 (1993), hep-th/9306083.
	
	\bibitem{AMPS}
	A. Almheiri, D. Marolf, J. Polchinski, and J. Sully, Black Holes: Complementarity or Firewalls?, JHEP \textbf{1302}, 062 (2013), arxiv: 1207.3123.
	
	\bibitem{HRT}
	V.E. Hubeny, M. Rangamani, T. Takayanagi, "A covariant holographic entanglement entropy proposal", JHEP \textbf{07}, 062 (2007), arxiv: 0705.0016
	
	\bibitem{H27}
	T. Faulkner, A. Lewkowycz, J. Maldacena,"Quantum corrections to holographic entanglement
	entropy," JHEP \textbf{11}, 074 (2013), arXiv:1307.2892.
	
	\bibitem{H56}
	N. Engelhardt, A. C. Wall, "Quantum Extremal Surfaces: Holographic Entanglement Entropy
	beyond the Classical Regime," JHEP \textbf{01}, 073 (2015),
	arXiv:1408.3203.
	
	
	\bibitem{Harlow}
	X. Dong, D. Harlow, A. C. Wall, "Reconstruction of bulk operators withing the entanglement wedge in gauge-gravity duality", Phys Rev Lett \textbf{117}, 021601 (2016), arxiv: 1601.05416.
	
	\bibitem{KM}
	I. Kourkoulou, J. Maldacena, "Pure states in the SYK model and nearly-AdS2 gravity", (2017), arXiv:1707.0232.
	
	\bibitem{P32}
	K. Papadodimas, S. Raju, "State-dependent bulk-boundary maps and black hole complementarity", Phys. Rev. D \textbf{89}, 086010 (2014), arxiv: 1310.6335.
	
	\bibitem{EWR_see_13}
	K. Papadodimas, S. Raju, "An Infalling Observer in AdS/CFT", JHEP \textbf{10}, 212 (2013), arxiv: 1211.6767.
	
	
	\bibitem{Almheiri2018}
	A. Almheiri, "Holographic Quantum Error Correction and the projected black hole interior", (2018), arxiv: 1810.02055
	
	
	\bibitem{Maldacena2017}
	J. Maldacena, D. Stanford, Z. Yang, "Diving into traversable wormholes",
	Fortsch.Phys. \textbf{65}, 1700034 (2017), arxiv: 1704.05333.
	
	\bibitem{EWR_see}
	A. Levine, A. Shabhazi-Moghaddam, R.M. Soni, "Seeing the entanglement wedge", (2020) arxiv: 2009.11305
	
	\bibitem{Bousso}
	arxiv: 2007.00230
	
	\bibitem{Witten}
	E. Witten,"Instability of the Kaluza-Klein vacuum", Nucl Phys B \textbf{195}, 481 (1982).
	
	\bibitem{Guillemin}
	B. Ammann, R. Lauter, V. Nistor, A. Vasy, "Complex powers and non-compact manifolds", Comm. Part.Diff. Eq. \textbf{29}, 5 (2002), arxiv: math/0211305.
	
	\bibitem{ComplexOp}
	D. Eelbode, "Arbitrary complex powers of the Dirac operator on the hyperbolic unit ball', Ann. Acad. Scient. Fenn. Math \textbf{29}, 367 (2004).
	
	\bibitem{Abramovitz}
	M. Abramowitz, I. A. Stegun, "Handbook of Mathematical Functions with Formulas, Graphs, and Mathematical Tables", Nat. Bur. Stand., App. Math Series \textbf{55} (1964).
	
	\bibitem{Visi}
	I.I. Cot\u{a}escu, M. Vi\c{s}inescu, "Symmetries and supersymmetries of the Dirac operators in curved spacetimes",  Progress in General Relativity and Quantum Cosmology Research, Nova Science, N.Y., 109 (2007), arxiv: hep-th/0411016. 
	
	\bibitem{Papadodimas}
	J. de Boer, R. van Breukelen, S.F. Lokhande, K. Papadodimas, E. Verlinde,
	"Probing typical black hole microstates", JHEP \textbf{01} (2020) 062, arxiv:
	1901.08527.
	
	\bibitem{CWitten}
	E. Witten, "A note on complex spacetime metrics", arxiv: 2111.06514.
	
	\bibitem{HongLiu}
	P. Gao, H. Liu, "Regenesis and quantum traversable wormholes", JHEP \textbf{10} (2019) 048, arxiv:1810.01444.
	
	\bibitem{TTWitten}
	E. Witten, "Notes on some entanglement properties of Quantum Field Theory", Rev. Mod. Phys. \textbf{90} (2018) 45003, arxiv: 1803.04993.
	
	\bibitem{Poland}
	R.F. Nalewajski, "Complex entropy and resultant information measures", J Math Chem \textbf{54} (2016) 1777.
	
	\bibitem{AlmheiriLin}
	A. Almheiri, H. Lin, "The entanglement wedge of unknown couplings", arxiv: 2111.06298.
	
\end{thebibliography}
\end{document}